\documentclass{article}
\usepackage{graphicx}
\usepackage{amsmath}
\usepackage{amsfonts}
\usepackage{amssymb}
\newtheorem{theorem}{Theorem}

\newtheorem{lemma}[theorem]{Lemma}

\newenvironment{proof}[1][Proof]{\textbf{#1.} }{\ \rule{0.5em}{0.5em}}

\begin{document}

\title{An analogue of the Sommerfeld radiation condition for the Dirac operator.}
\author{Vladislav V. Kravchenko, Ra\'{u}l Castillo P.\\Departamento de Telecomunicaciones,\\Escuela Superior de Ingenier\'{i}a Mec\'{a}nica y El\'{e}ctrica,\\Instituto Polit\'{e}cnico Nacional, C.P. 07738, D.F., MEXICO\\e-mail: vkravche@maya.esimez.ipn.mx\\\ \ \ \ \ \ \ \ \ \ raulcp@tesla.esimez.ipn.mx}
\date{}
\maketitle
\begin{abstract}
A simple radiation condition at infinity for time-harmonic massive Dirac
spinors is proposed. This condition allows an analogue of the Cauchy integral
formula in unbounded domains for null-solutions of the Dirac equation to be
proved. The result is obtained with the aid of methods of quaternionic analysis.
\end{abstract}

\textbf{Key words: }Dirac operator, radiation condition.\medskip

\textbf{AMS subject classification:} 30G35, 81Q05.\medskip

\section{\textbf{Introduction}\medskip}

There are two principal ways to guarantee the unicity of solution of a not
homogeneous differential equation of mathematical physics in the whole space.
The first is the famous limiting absorption principle which in the case of
some operators, for example the Helmholtz operator, has a very natural
physical meaning reflected in its name (see, e.g., \cite{EOMP}). The medium is
assumed to be slightly absorbent with a corresponding absorption parameter
$\varepsilon.$ For $\varepsilon>0$ the inhomogeneous equation (under some
additional natural conditions) has a unique solution $u_{\varepsilon}$
decreasing at infinity, and the solution for $\varepsilon=0$ is obtained as
the limit of $u_{\varepsilon}$ when $\varepsilon\rightarrow0.$ This approach,
based on a deep understanding of the physical nature of the differential
operator, is not applicable in all cases. Another approach is to impose a
radiation condition at infinity which ``prohibits'' the existence of some
unwanted solutions. Such a prohibition is related to the unicity of an
integral representation for null-solutions of the differential operator. In
the case of the Helmholtz operator the corresponding radiation condition was
introduced by A. Sommerfeld \cite{Sommer}.\medskip

For the Maxwell equations such condition was proposed by C. M\"{u}ller
\cite{Muller} and S. Silver \cite{Silver}. Recently, A. McIntosh and M. Mitrea
in \cite{CAME} introduced a radiation condition for a special perturbed
Clifford-Dirac operator. We use their approach to obtain a radiation condition
for the operator $D+M^{\overrightarrow{\alpha}}$, where $D$ is the
Moisil-Theodoresco operator and $M^{\overrightarrow{\alpha}}$ is the operator
of multiplication by a complex quaternion $\overrightarrow{\alpha}$ from the
right-hand side (for exact definitions see Section 2). Then using a simple
matrix transformation proposed in \cite{OBBM}, which converts the operator
$D+M^{\overrightarrow{\alpha}}$ into the Dirac operator in its traditional
form through the $\gamma$-matrices, we rewrite the radiation condition for the
Dirac spinors in a form suitable for applications and obtain an integral
representation for null-solutions of the classical Dirac operator in unbounded
domains, based on the proposed radiation condition.\bigskip

\section{\textbf{Preliminaries}}

\subsection{\bigskip\textbf{Quaternions}}

A quaternion $a$\ can be represented as a linear combination of its components
($\{a_{k}\}\subset\mathbb{R}$ for real quaternions and $\{a_{k}\}\subset
\mathbb{C}$ for complex quaternions) and the elements of the orthonormal basis
$i_{k}$ as follows
\[
a=\sum_{k=0}^{3}a_{k}i_{k}%
\]
where $i_{0}:=1$ and $\{i_{k}\mid k=1,2,3\}$ are the quaternionic imaginary
units, which have the following properties:
\[
i_{k}^{2}=-1,\quad\qquad k=1,2,3;
\]%
\begin{equation}
i_{1}i_{2}=-i_{2}i_{1}=i_{3};\quad i_{2}i_{3}=-i_{3}i_{2}=i_{1};\quad
\ i_{3}i_{1}=-i_{1}i_{3}=i_{2}.
\end{equation}

The imaginary unit in $\mathbb{C}$ is denoted as usual by $i$ and by
definition commutes with quaternionic imaginary units.
\[
i\cdot i_{k}=i_{k}\cdot i,\qquad k=\overline{0,3}.
\]

In some cases it is useful to represent a quaternion $a$ as the sum of a
scalar part $\operatorname{Sc}(a)=a_{0}$ and a vectorial part
$\operatorname*{Vec}(a)=\overrightarrow{a}=a_{1}i_{1}+a_{2}i_{2}+a_{3}i_{3}$.
Using this representation we can define the product of two arbitrary
quaternions $a$ and $b$ in the form
\begin{equation}
a\cdot b=a_{0}b_{0}-\left\langle \overrightarrow{a},\overrightarrow
{b}\right\rangle +a_{0}\overrightarrow{b}+\overrightarrow{a}b_{0}+\left[
\overrightarrow{a}\times\overrightarrow{b}\right]  , \label{qproduct}%
\end{equation}
where $\left\langle \cdot,\cdot\right\rangle $ is the scalar product and
$\left[  \cdot\times\cdot\right]  $ is the vector product.\medskip

The quaternionic conjugation of a quaternion $a=a_{0}+\overrightarrow{a}$ is
defined as $\overline{a}=a_{0}-\overrightarrow{a}.$ From (\ref{qproduct}) it
is easy to observe that
\begin{equation}
a\cdot\overline{a}=a_{0}^{2}+a_{1}^{2}+a_{2}^{2}+a_{3}^{2}=:\left|  a\right|
^{2}, \label{qbyconj}%
\end{equation}
and to obtain an important property of quaternionic conjugation:
\[
\overline{a\cdot b}=\overline{b}\cdot\overline{a}%
\]
Note that $\left|  a\cdot b\right|  =\left|  a\right|  \cdot\left|  b\right|
.$\medskip

One important conclusion which can be made using (\ref{qbyconj}) is that each
non-zero real quaternion $a$ is invertible and its inverse is given by
\[
a^{-1}=\frac{\overline{a}}{\left|  a\right|  ^{2}}.
\]

We denote the set of zero divisors in $\mathbb{H(C)}$ by $\frak{S}$:
\[
\frak{S}:=\{a\in\mathbb{H}(\mathbb{C})\mid a\neq0;\exists b\neq0:a\cdot
b=0\}.
\]

\bigskip

\begin{lemma}
[\cite{IRSM}](Structure of the set of zero divisors) Let $a\in\mathbb{H(C)}$
and $a\neq0.$ The following statements are equivalent:

1. $a\in\frak{S}$.

2. $a\cdot\overline{a}=0.$

3. $a_{0}^{2}=\overrightarrow{a}^{2}.$

4. $a^{2}=2a_{0}a=2\overrightarrow{a}a.$
\end{lemma}

As the modulus introduced by (\ref{qbyconj}) does not in general give
information about the absolute values of the components of a complex
quaternion when it is a zero divisor, another kind of modulus is frequently
used
\begin{equation}
\left|  a\right|  _{c}:=\sqrt{\left|  a_{0}\right|  ^{2}+\left|  a_{1}\right|
^{2}+\left|  a_{2}\right|  ^{2}+\left|  a_{3}\right|  ^{2}}, \label{modulecq}%
\end{equation}
where $\left|  a_{k}\right|  ^{2}=a_{k}a_{k}^{\ast}$ and ``*'' stands for the
usual complex conjugation. (\ref{modulecq}) can be expressed also in the
following ways
\[
\left|  a\right|  _{c}^{2}=\left|  \operatorname{Re}a\right|  ^{2}+\left|
\operatorname{Im}a\right|  ^{2}=\operatorname{Sc}(a\cdot\overline{a}^{\ast
})=\operatorname{Sc}(\overline{a}^{\ast}\cdot a).
\]
\medskip

\subsection{Quaternionic differential operators\bigskip}

Let $\Omega\subset\mathbb{R}^{3}$ and $f\in C^{1}(\Omega;\mathbb{H}%
(\mathbb{C}))$. The Moisil-Theodoresco operator is defined as
\[
Df:=\sum_{k=1}^{3}i_{k}\partial_{k}f
\]
where $\partial_{k}:=\frac{\partial}{\partial x_{k}}.$ An important property
of $D$ is obtained from (\ref{qbyconj})
\begin{equation}
D^{2}=-\Delta, \label{delta}%
\end{equation}
where $\Delta=\partial_{1}^{2}+\partial_{2}^{2}+\partial_{3}^{2}$ is the usual
Laplace operator.\bigskip

We will consider the following operator
\[
D_{\pm\nu}:=D\pm\nu I
\]
where $\nu$ is a complex number, which was studied for the first time in
\cite{HFHE} (for $\nu$ real). With the aid of the operator $D_{\nu}$ it is
possible to factorize the Helmholtz operator
\begin{equation}
\Delta+\nu^{2}=-(D+\nu)(D-\nu)=-D_{\nu}D_{-\nu}. \label{facthelm}%
\end{equation}

Suppose that $\theta_{\nu}$ is a fundamental solution of the Helmholtz
operator
\[
(\Delta+\nu^{2})\theta_{\nu}=\delta.
\]
Then, using (\ref{facthelm}) we obtain the functions
\begin{equation}
\mathcal{K}_{\pm\nu}:=-(D\mp\nu)\theta_{\nu} \label{fundsolda}%
\end{equation}
which are the fundamental solutions of $D_{\pm\nu}$, that is
\[
D_{\pm\nu}\mathcal{K}_{\pm\nu}=\delta.
\]

Let us assume that $\operatorname{Im}\nu\geq0.$ Under this condition, the
fundamental solution of the Helmholtz operator is chosen as
\begin{equation}
\theta_{\nu}(x)=-\frac{e^{i\nu\left|  x\right|  }}{4\pi\left|  x\right|  },
\label{fundsolhelm}%
\end{equation}
in which case it fulfills the Sommerfeld radiation condition at infinity and
has a clear physical sense of an outgoing wave generated by a point source at
the origin.\medskip

Substituting (\ref{fundsolhelm}) into equality (\ref{fundsolda}), we obtain
\[
\mathcal{K}_{\pm\nu}(x)=-D\theta_{\nu}(x)\pm\nu\theta_{\nu}(x)=\left(  \pm
\nu+\frac{x}{\left|  x\right|  ^{2}}-i\nu\frac{x}{\left|  x\right|  }\right)
\theta_{\nu}(x),
\]
where $x:=\sum_{k=1}^{3}x_{k}i_{k}.$ These functions were obtained in
\cite{ORBH}, (see also \cite[Section 3]{IRSM}).\medskip

Let us assume that $\Omega$ is a bounded domain in $\mathbb{R}^{3}$ with a
piecewise smooth boundary $\Gamma:=\partial\Omega$.\medskip

\begin{theorem}
\label{Stok}(Quaternionic Stokes' formula) Let $f$ and $g$ belong to
$C^{1}(\Omega;\mathbb{H}(\mathbb{C}))\cap C(\overline{\Omega};\mathbb{H}%
(\mathbb{C}))$. Then
\begin{equation}
\int_{\Omega}\left(  (D_{r}f(y))g(y)+f(y)(Dg(y))\right)  dy=\int_{\Gamma
}f(y)\overrightarrow{n}(y)g(y)d\Gamma_{y}, \label{Stokes}%
\end{equation}
where $\overrightarrow{n}(y)$ is the quaternionic representation of the
outward unit normal to the surface $\Gamma:\overrightarrow{n}(y)=\sum
\nolimits_{k=1}^{3}i_{k}n_{k}(y)$ and $D_{r}f:=\sum_{k=1}^{3}\partial
_{k}fi_{k}$ is the right Moisil-Theodoresco operator.\medskip
\end{theorem}

The proof of this fact \ can be found in \cite[Chapter 4]{IRSM} or \cite[p.
86]{QCCP}.\medskip

Let us consider the following integral operator
\begin{equation}
K_{\nu}[f](x):=-\int_{\Gamma}\mathcal{K}_{\nu}(x-y)\overrightarrow
{n}(y)f(y)d\Gamma_{y},\quad x\in\mathbb{R}^{3}\setminus\Gamma\label{K}%
\end{equation}
\medskip which can be considered as an analogue of the Cauchy integral
operator from complex analysis due to the following fact.\medskip

\begin{theorem}
\label{Cauchy}(Quaternionic Cauchy integral formula) Let $f\in C^{1}%
(\Omega;\mathbb{H}(\mathbb{C}))\cap C(\overline{\Omega};\mathbb{H}%
(\mathbb{C}))$ and $f\in\ker D_{\nu}(\Omega)$. Then
\begin{equation}
f(x)=K_{\nu}[f](x),\quad\forall x\in\Omega.\label{Cauchyf}%
\end{equation}
\medskip
\end{theorem}

\subsection{\textbf{Dirac operator in quaternionic form}\medskip}

Denote $\widetilde{\Phi}:=\Phi(t,x_{1},x_{2},-x_{3}).$ The domain
$\widetilde{\mathcal{G}}$ is assumed to be obtained from the domain
$\mathcal{G}\subset\mathbb{R}^{4}$ by the reflection $x_{3}\rightarrow-x_{3}.$
Let us consider the transformation denoted by $A$ and introduced in the
following way. A function $\Phi:\mathcal{G}\subset\mathbb{R}^{4}%
\rightarrow\mathbb{C}^{4}$ is transformed into a function $\rho$
$:\widetilde{\mathcal{G}}\subset\mathbb{R}^{4}\rightarrow\mathbb{H}%
(\mathbb{C})$ by the rule
\[
\rho=A[\Phi]:=\frac{1}{2}\left(  -(\widetilde{\Phi_{1}}-\widetilde{\Phi_{2}%
})i_{0}+i(\widetilde{\Phi_{0}}-\widetilde{\Phi_{3}})i_{1}-(\widetilde{\Phi
_{0}}+\widetilde{\Phi_{3}})i_{2}+i(\widetilde{\Phi_{1}}+\widetilde{\Phi_{2}%
})i_{3}\right)  .
\]
Then the definition of the inverse transformation $A^{-1}$ is
\[
A^{-1}[\rho]=(-i\widetilde{\rho}_{1}-\widetilde{\rho}_{2},-\widetilde{\rho
}_{0}-i\widetilde{\rho}_{3},\widetilde{\rho}_{0}-i\widetilde{\rho}%
_{3},i\widetilde{\rho}_{1}-\widetilde{\rho}_{2}).
\]

These transformations can be represented in a more explicit matrix form:
\[
\rho=A[\Phi]=\frac{1}{2}\left(
\begin{array}
[c]{rrrr}%
0 & -1 & 1 & 0\\
i & 0 & 0 & -i\\
-1 & 0 & 0 & -1\\
0 & i & i & 0
\end{array}
\right)  \left(
\begin{array}
[c]{c}%
\widetilde{\Phi_{0}}\\
\widetilde{\Phi_{1}}\\
\widetilde{\Phi_{2}}\\
\widetilde{\Phi_{3}}%
\end{array}
\right)
\]
and
\[
\Phi=A^{-1}[\rho]=\left(
\begin{array}
[c]{rrrr}%
0 & -i & -1 & 0\\
-1 & 0 & 0 & -i\\
1 & 0 & 0 & -i\\
0 & i & -1 & 0
\end{array}
\right)  \left(
\begin{array}
[c]{c}%
\widetilde{\rho}_{0}\\
\widetilde{\rho}_{1}\\
\widetilde{\rho}_{2}\\
\widetilde{\rho}_{3}%
\end{array}
\right)  .
\]

The Dirac $\gamma$-matrices in the standard (\cite{QF}, \cite{TDE})
Dirac-Pauli form are
\[%
\begin{array}
[c]{cc}%
\gamma_{0}:=\left(
\begin{array}
[c]{rrrr}%
1 & 0 & 0 & 0\\
0 & 1 & 0 & 0\\
0 & 0 & -1 & 0\\
0 & 0 & 0 & -1
\end{array}
\right)  ,\qquad & \gamma_{1}:=\left(
\begin{array}
[c]{rrrr}%
0 & 0 & 0 & -1\\
0 & 0 & -1 & 0\\
0 & 1 & 0 & 0\\
1 & 0 & 0 & 0
\end{array}
\right)  ,
\end{array}
\]%
\[%
\begin{array}
[c]{cc}%
\gamma_{2}:=\left(
\begin{array}
[c]{rrrr}%
0 & 0 & 0 & i\\
0 & 0 & -i & 0\\
0 & -i & 0 & 0\\
i & 0 & 0 & 0
\end{array}
\right)  ,\qquad & \gamma_{3}:=\left(
\begin{array}
[c]{rrrr}%
0 & 0 & -1 & 0\\
0 & 0 & 0 & 1\\
1 & 0 & 0 & 0\\
0 & -1 & 0 & 0
\end{array}
\right)  ;
\end{array}
\]%
\[
\gamma_{5}:=i\gamma_{0}\gamma_{1}\gamma_{2}\gamma_{3}=\left(
\begin{array}
[c]{rrrr}%
0 & 0 & -1 & 0\\
0 & 0 & 0 & -1\\
-1 & 0 & 0 & 0\\
0 & -1 & 0 & 0
\end{array}
\right)  .
\]

Recall that:

1. $\gamma_{0}^{2}=\gamma_{5}^{2}=E_{4},$ the identity matrix;

2. $\gamma_{k}^{2}=-E_{4},$ $k=1,2,3$;

3. $\gamma_{j}\gamma_{k}=-\gamma_{k}\gamma_{j}$, $j,k=0,1,2,3,5,$ $j\neq k$.\medskip

The transforms $A$ and $A^{-1}$ have the following algebraic properties
\cite{OBBM}:

1. $A\gamma_{1}\gamma_{2}\gamma_{3}\gamma_{1}A^{-1}[\rho]=i_{1}\rho;$

2. $A\gamma_{1}\gamma_{2}\gamma_{3}\gamma_{2}A^{-1}[\rho]=i_{2}\rho;$

3. $A\gamma_{1}\gamma_{2}\gamma_{3}\gamma_{3}A^{-1}[\rho]=-i_{3}\rho;$

4. $A\gamma_{1}\gamma_{2}\gamma_{3}\gamma_{0}A^{-1}[\rho]=\rho i_{1};$

5. $A\gamma_{1}\gamma_{2}\gamma_{3}A^{-1}[\rho]=-i\rho i_{2}.$\medskip

The Dirac operator for a free massive particle of spin 1/2 is
\[
\mathbb{D}[\Phi]:=\left(  \gamma_{0}\partial_{t}-\sum_{k=1}^{3}\gamma
_{k}\partial_{k}+im\right)  [\Phi],
\]
where $m\in\mathbb{R}$ and $\Phi$ is a $\mathbb{C}^{4}$-valued function.
Consider the quaternionic differential operator
\[
R:=P_{1}^{+}(i\partial_{t}+D)+P_{1}^{-}(-i\partial_{t}+D)-mM^{i_{2}},
\]
where $P_{k}^{\pm}:=\frac{1}{2}M^{(1\pm ii_{k})}$ and $M^{\lambda
}[f]:=f\lambda,$ $\lambda\in\mathbb{H}(\mathbb{C})$. Using the algebraic
properties of $A$ and $A^{-1}$ we obtain the following equality \cite{OBBM}:
\begin{equation}
R=-A\gamma_{1}\gamma_{2}\gamma_{3}\mathbb{D}A^{-1}. \label{diracmatrix}%
\end{equation}

We consider a time-harmonic null-solution $\Phi$ of the operator $\mathbb{D} $
which has the form $\Phi(t,x)=q(x)e^{i\omega t}$, where $\omega\in\mathbb{R}$
and $q(x)$ is a $\mathbb{C}^{4}$-valued function depending only on $x$. We
have the following equation for $q$%
\[
\mathbb{D}_{\omega,m}q(x)=0,
\]
where $\mathbb{D}_{\omega,m}:=i\omega\gamma_{0}-\sum_{k=1}^{3}\gamma
_{k}\partial_{k}+imI$. The corresponding quaternionic reformulation of
$\mathbb{D}_{\omega,m}$ is the operator
\[
D_{\alpha}:=D+M^{\alpha},
\]
where $\alpha:=-(i\omega i_{1}+mi_{2}).$ That is
\begin{equation}
D_{\alpha}=-A\gamma_{1}\gamma_{2}\gamma_{3}\mathbb{D}_{\omega,m}%
A^{-1}.\medskip\label{Dc}%
\end{equation}

\section{Radiation condition for the Dirac operator\medskip}

In order to prove the Cauchy integral formula for null-solutions of the
operator $D_{\nu}$ for the exterior domain $\mathbb{R}^{3}\backslash
\overline{\Omega}$ we need some appropriate radiation condition at infinity.\medskip

Let us consider the equation
\[
D_{\nu}\mathcal{K}(x)=\delta(x),\quad\nu\neq0,\quad x\in\mathbb{R}^{3}.
\]
When $\operatorname{Im}\nu=0$ it admits two solutions decreasing at infinity
obtained (according to (\ref{fundsolda})) by the application of the operator
$-D_{-\nu}$ to the fundamental solutions of the Helmholtz operator $u^{\pm
}(x):=-\frac{e^{\pm i\nu\left|  x\right|  }}{4\pi\left|  x\right|  }$
respectively:
\[
\mathcal{K}^{\pm}(x)=\left(  \nu+\frac{x}{\left|  x\right|  ^{2}}\mp i\nu
\frac{x}{\left|  x\right|  }\right)  \cdot u^{\pm}(x).
\]
In order to omit one of these possibilities we impose the following radiation
condition
\begin{equation}
\left(  \nu-\frac{x}{\left|  x\right|  ^{2}}+i\nu\frac{x}{\left|  x\right|
}\right)  \cdot\mathcal{K}(x)=o\left(  \frac{1}{\left|  x\right|  }\right)
,\quad\text{when\quad}\left|  x\right|  \rightarrow\infty. \label{rad}%
\end{equation}
\medskip It is easy to see that
\[
\left(  \nu-\frac{x}{\left|  x\right|  ^{2}}+i\nu\frac{x}{\left|  x\right|
}\right)  \mathcal{K}^{+}(x)=O\left(  \frac{1}{\left|  x\right|  ^{2}}\right)
,\quad\text{when }\left|  x\right|  \rightarrow\infty
\]
and hence $\mathcal{K}^{+}$ fulfills (\ref{rad}). For the function
$\mathcal{K}^{-}$ this is not true. Note that $\mathcal{K}^{+}$ is precisely
the fundamental solution $\mathcal{K}_{\nu}$.\medskip

We are now ready to prove the following \medskip

\begin{theorem}
\label{Cauchyextt}(Quaternionic Cauchy integral formula for the exterior
domain) Let
\[
f\in C^{1}(\mathbb{R}^{3}\setminus\overline{\Omega};\mathbb{H}(\mathbb{C}%
))\cap C(\mathbb{R}^{3}\setminus\Omega;\mathbb{H}(\mathbb{C})),\quad f\in\ker
D_{\nu}(\mathbb{R}^{3}\setminus\overline{\Omega}),\quad\operatorname*{Im}%
\nu\geq0
\]
and let $f$ satisfy the radiation condition
\[
\left(  \nu-\frac{x}{\left|  x\right|  ^{2}}+i\nu\frac{x}{\left|  x\right|
}\right)  \cdot f(x)=o\left(  \frac{1}{\left|  x\right|  }\right)
,\quad\text{when}\quad\left|  x\right|  \rightarrow\infty.
\]
Then
\begin{equation}
f(x)=-K_{\nu}[f](x),\quad\forall x\in\mathbb{R}^{3}\setminus\overline{\Omega
}.\label{Cauchyext}%
\end{equation}
\end{theorem}

\begin{proof}
Let $\Gamma^{R}$ be a sphere with center at the origin and radius $R$
sufficiently large so that $\Omega$ is contained in the ball $B^{R}$ with
boundary $\partial B^{R}=\Gamma^{R}$ (Fig. \ref{2}).%
\begin{figure}
[ptb]
\begin{center}
\includegraphics[
natheight=2.009000in,
natwidth=2.250200in,
height=1.9943in,
width=2.2303in
]%
{Fig2.bmp}%
\label{2}%
\end{center}
\end{figure}

According to Theorem \ref{Cauchy} in each point $x$ of the domain $\Omega
^{R}:=B^{R}\setminus\overline{\Omega}$ we have the equality
\[
f(x)=\int_{\Gamma}\mathcal{K}_{\nu}(x-y)\overrightarrow{n}(y)f(y)d\Gamma
_{y}-\int_{\Gamma^{R}}\mathcal{K}_{\nu}(x-y)\frac{y}{\left|  y\right|
}f(y)d\Gamma_{y}^{R}.
\]
We now consider the limit of this equality when $R\rightarrow\infty$. We have
the following asymptotic relation
\[
\int_{\Gamma^{R}}\mathcal{K}_{\nu}(x-y)\frac{y}{\left|  y\right|  }%
f(y)d\Gamma_{y}^{R}\sim\int_{\Gamma^{R}}\theta(y)\left(  \nu-\frac{y}{\left|
y\right|  ^{2}}+i\nu\frac{y}{\left|  y\right|  }\right)  f(y)d\Gamma_{y}%
^{R},\quad R\rightarrow\infty.
\]
Using the radiation condition we obtain that this integral tends to zero when
$R\rightarrow\infty$. Thus,
\[
f(x)=\int_{\Gamma}\mathcal{K}_{\nu}(x-y)\overrightarrow{n}(y)f(y)d\Gamma_{y}.
\]
\end{proof}

The introduction of the radiation condition in the form (\ref{rad}) allowed us
to obtain a very simple proof of the Cauchy integral formula for the exterior
domain. Nevertheless, as was shown in \cite{CAME} this condition can be
represented in a more elegant form. We note that the expression which appears
in (\ref{rad}) can be rewritten as follows
\[
\left(  \nu-\frac{x}{\left|  x\right|  ^{2}}+i\nu\frac{x}{\left|  x\right|
}\right)  =\nu\left(  1+\frac{ix}{\left|  x\right|  }\right)  +O\left(
\frac{1}{\left|  x\right|  }\right)  ,\quad\left|  x\right|  \rightarrow
\infty.
\]
Thus, a natural idea is to introduce the radiation condition as
\begin{equation}
\left(  1+\frac{ix}{\left|  x\right|  }\right)  f(x)=o\left(  \frac{1}{\left|
x\right|  }\right)  , \label{radMM}%
\end{equation}
because the term $x/\left|  x\right|  ^{2}$ apparently gives a faster decay.
The problem here is that (\ref{radMM}) does not imply the decay of the product
$\frac{x}{\left|  x\right|  ^{2}}\cdot f(x)$ when $\left|  x\right|
\rightarrow\infty$ due to the fact that $(1+\frac{ix}{\left|  x\right|  })$ is
a zero divisor. The following example makes this clear. Consider
$f(x):=(1-\frac{ix}{\left|  x\right|  })\cdot\left|  x\right|  ^{2}$. This
function obviously satisfies (\ref{radMM}) but $\frac{x}{\left|  x\right|
^{2}}\cdot f(x)=O(\left|  x\right|  )$, $\left|  x\right|  \rightarrow\infty$.
Thus, in order to be able to apply (\ref{radMM}) instead of (\ref{rad}), we
should prove that if $f$ belongs to $\ker D_{\nu}(\mathbb{R}^{3}%
\setminus\overline{\Omega})$ and satisfies (\ref{radMM}) then it decreases at
infinity. Following arguments from \cite{CAME} we prove this.\medskip

\begin{theorem}
Let $f\in C^{1}(\mathbb{R}^{3}\setminus\overline{\Omega};\mathbb{H}%
(\mathbb{C}))\cap C(\mathbb{R}^{3}\setminus\Omega;\mathbb{H}(\mathbb{C}%
)),\quad f\in\ker D_{\nu}(\mathbb{R}^{3}\setminus\overline{\Omega}%
),\quad\operatorname*{Im}\nu\geq0$, such that satisfies (\ref{radMM}). Then
\begin{equation}
\int_{\left|  x\right|  =R}\left|  f(x)\right|  _{c}^{2}d\Gamma_{x}%
^{R}=O(1)\quad\text{as\quad}R\rightarrow\infty. \label{estim}%
\end{equation}
\medskip
\end{theorem}

\begin{proof}
Consider the expression
\begin{align*}
\left|  \left(  1+\frac{ix}{\left|  x\right|  }\right)  f\right|  _{c}^{2}  &
=\operatorname{Sc}\left(  \left(  1+\frac{ix}{\left|  x\right|  }\right)
f\overline{f}^{\ast}\left(  1+\frac{ix}{\left|  x\right|  }\right)  \right)
=2\operatorname*{Sc}\left(  \left(  1+\frac{ix}{\left|  x\right|  }\right)
f\overline{f}^{\ast}\right)  =\\
& \\
&  =2\left(  \left|  f\right|  _{c}^{2}-\operatorname*{Im}\operatorname*{Sc}%
(\overline{f}^{\ast}\frac{x}{\left|  x\right|  }f)\right)  .
\end{align*}
Consequently,
\begin{align}
\int_{\left|  x\right|  =R}\left|  \left(  1+\frac{ix}{\left|  x\right|
}\right)  f(x)\right|  _{c}^{2}d\Gamma_{x}^{R}  &  =2(\int_{\left|  x\right|
=R}\left|  f(x)\right|  _{c}^{2}d\Gamma_{x}^{R}-\label{vsp2}\\
& \nonumber\\
&  \operatorname*{Im}\operatorname*{Sc}(\int_{\left|  x\right|  =R}%
\overline{f}^{\ast}(x)\frac{x}{\left|  x\right|  }f(x)d\Gamma_{x}%
^{R})).\nonumber
\end{align}

Due to Theorem \ref{Stok} we have (using the notation of Theorem
\ref{Cauchyextt})
\begin{align*}
\int_{\left|  x\right|  =R}\overline{f}^{\ast}(x)\frac{x}{\left|  x\right|
}f(x)d\Gamma_{x}^{R}  &  =\int_{\Gamma}\overline{f}^{\ast}(x)\overrightarrow
{n}(x)f(x)d\Gamma_{x}+\\
& \\
&  \int_{\Omega^{R}}(D_{r}(\overline{f}^{\ast}(x))\cdot f(x)+\overline
{f}^{\ast}(x)\cdot Df(x))dx.
\end{align*}
The first integral on the right-hand side is some constant $C$. In order to
simplify the second we use $Df=-\nu f$ and observe that
\[
D_{r}\overline{f}^{\ast}=-\overline{Df^{\ast}}=\overline{\nu^{\ast}f^{\ast}%
}=\nu^{\ast}\overline{f}^{\ast}.
\]
Then
\begin{align*}
\int_{\Omega^{R}}(D_{r}(\overline{f}^{\ast}(x))\cdot f(x)+\overline{f}^{\ast
}(x)\cdot Df(x))dx  &  =\int_{\Omega^{R}}(\nu^{\ast}\overline{f}^{\ast
}(x)\cdot f(x)-\nu\overline{f}^{\ast}(x)\cdot f(x))dx=\\
& \\
&  =-2i\operatorname*{Im}\nu\int_{\Omega^{R}}\overline{f}^{\ast}(x)\cdot
f(x)dx,
\end{align*}
we obtain that
\[
\operatorname*{Sc}\left(  \int_{\left|  x\right|  =R}\overline{f}^{\ast
}(x)\frac{x}{\left|  x\right|  }f(x)d\Gamma_{x}^{R}\right)  =C_{0}%
-2i\operatorname*{Im}\nu\int_{\Omega^{R}}\left|  f(x)\right|  _{c}^{2}dx.
\]
Substituting this expression in (\ref{vsp2}) and using the radiation condition
(\ref{radMM}) we have
\[
\int_{\left|  x\right|  =R}\left|  f(x)\right|  _{c}^{2}d\Gamma_{x}%
^{R}=\operatorname*{Im}C_{0}-2\operatorname*{Im}\nu\int_{\Omega^{R}}\left|
f(x)\right|  _{c}^{2}dx,\quad R\rightarrow\infty.
\]
The last term above vanishes when $\operatorname*{Im}\nu=0$ and is negative
when $\operatorname*{Im}\nu>0$. In both cases (\ref{estim}) is proved.\medskip
\end{proof}

This theorem establishes the equivalence of the radiation conditions
(\ref{rad}) and (\ref{radMM}) for the functions from $\ker D_{\nu}$.\medskip

We now obtain the radiation condition for the Dirac operator in quaternionic
form, that is for $D_{\alpha}$, and prove the Cauchy integral formula for the
exterior domain for this operator. After that, using the transform $A$ we will
rewrite these results for the operator $\mathbb{D}_{\omega,m}.$\medskip

Let $\alpha$ be a purely vectorial complex quaternion $\alpha=\overrightarrow
{\alpha}$, $\alpha\notin\frak{S}$, $\overrightarrow{\alpha}^{2}\neq0$ and
consider $\nu:=\sqrt{\overrightarrow{\alpha}^{2}}\in\mathbb{C}$ where the sign
of the root is chosen in a way such that $\operatorname*{Im}$ $\nu\geq0.$\medskip

Let
\begin{equation}
K_{\overrightarrow{\alpha}}[f](x):=P^{+}K_{\nu}[f](x)+P^{-}K_{-\nu}[f](x)
\end{equation}
where the operators $K_{\pm\nu}$ are defined by (\ref{K}) and $P^{\pm}%
:=\frac{1}{2\nu}M^{(\nu\pm\overrightarrow{\alpha})}$ are two mutually
complementary projection operators in the set of $\mathbb{H(C)}$-valued functions.\medskip

In the case when $\alpha:=\overrightarrow{\alpha}\in\frak{S}$, $\nu=0,$ we
introduce the operator $K_{\overrightarrow{\alpha}}$ as follows
\begin{equation}
K_{\overrightarrow{\alpha}}f:=K_{0}f-M^{\overrightarrow{\alpha}}V_{0}f
\label{kalfa}%
\end{equation}
where
\begin{equation}
V_{0}[f](x):=\int\limits_{\Gamma}\theta_{0}(x-y)\overrightarrow{n}%
(y)f(y)d\Gamma_{y},\qquad x\in\mathbb{R}^{3}\setminus\Gamma. \label{V0}%
\end{equation}
Here $\theta_{0}(x-y)=-\frac{1}{4\pi\left|  x-y\right|  }.$\medskip

For all $\alpha\in\mathbb{H(C)}$ (that is $\alpha\in\frak{S}$ or $\alpha
\notin\frak{S}$), $K_{\overrightarrow{\alpha}}f$ admits the following
representation
\begin{multline}
K_{\overrightarrow{\alpha}}[f](x)=-\int\limits_{\Gamma}\{\theta_{\nu
}(x-y)\left(  \frac{x-y}{\left|  x-y\right|  ^{2}}-\frac{i\nu(x-y)}{\left|
x-y\right|  }\right)  \overrightarrow{n}(y)f(y)+\\
\nonumber\\
+\theta_{\nu}(x-y)\overrightarrow{n}(y)f(y)\overrightarrow{\alpha}%
\}d\Gamma_{y},\qquad x\in\Gamma.\nonumber
\end{multline}
\medskip

\begin{theorem}
Let $f\in C^{1}(\mathbb{R}^{3}\setminus\overline{\Omega};\mathbb{H}%
(\mathbb{C}))\cap C(\mathbb{R}^{3}\setminus\Omega;\mathbb{H}(\mathbb{C}))$,
$f\in\ker D_{\alpha}(\mathbb{R}^{3}\setminus\overline{\Omega})$, $\alpha
\in\mathbb{H(C)}$, $\operatorname{Sc}\alpha=0$ such that $f$ satisfies the
radiation condition
\begin{equation}
\nu f(x)+\frac{ix}{\left|  x\right|  }f(x)\overrightarrow{\alpha}=o\left(
\frac{1}{\left|  x\right|  }\right)  ,\qquad\text{when }\left|  x\right|
\rightarrow\infty, \label{radcondition}%
\end{equation}
where $\nu:=\sqrt{\overrightarrow{\alpha}^{2}}\in\mathbb{C}$ and
$\operatorname{Im}\nu\geq0.$ If $\alpha\in\frak{S}$ then we suppose
additionally that $f(x)=o(1).$ Then
\[
f(x)=K_{\overrightarrow{\alpha}}[f](x),\qquad\forall x\in\mathbb{R}%
^{3}\setminus\overline{\Omega}.
\]
\medskip
\end{theorem}

\begin{proof}
First we suppose that $\overrightarrow{\alpha}\notin\frak{S}$. Multiplying
(\ref{radcondition}) by $\overrightarrow{\alpha}/\nu$ from the right-hand side
we obtain
\begin{equation}
f(x)\overrightarrow{\alpha}+\nu\frac{ix}{\left|  x\right|  }f(x)=o\left(
\frac{1}{\left|  x\right|  }\right)  \label{2ndradcond}%
\end{equation}
adding and subtracting (\ref{radcondition}) and (\ref{2ndradcond}) we obtain
\[
\left(  \nu f(x)+\frac{ix}{\left|  x\right|  }f(x)\overrightarrow{\alpha
}\right)  \pm\left(  f(x)\overrightarrow{\alpha}+\nu\frac{ix}{\left|
x\right|  }f(x)\right)  =
\]%
\[
=\left(  1\pm\frac{ix}{\left|  x\right|  }\right)  f(x)\nu\pm\left(  1\pm
\frac{ix}{\left|  x\right|  }\right)  f(x)\overrightarrow{\alpha}=
\]%
\begin{equation}
=\left(  1\pm\frac{ix}{\left|  x\right|  }\right)  f(x)\left(  \nu
\pm\overrightarrow{\alpha}\right)  =o\left(  \frac{1}{\left|  x\right|
}\right)  . \label{radprojector}%
\end{equation}
Which can be writen as follows
\begin{equation}
P^{+}\left(  \left(  1+\frac{ix}{\left|  x\right|  }\right)  f(x)\right)
+P^{-}\left(  \left(  1-\frac{ix}{\left|  x\right|  }\right)  f(x)\right)
=o\left(  \frac{1}{\left|  x\right|  }\right)  \quad\text{when }\left|
x\right|  \rightarrow\infty. \label{projections}%
\end{equation}
Thus, (\ref{radcondition}) is equivalent to (\ref{projections}), from which it
can be seen that $P^{+}f$ fulfills the radiation condition (\ref{radMM}) and
$P^{-}f$ fulfills its conjugate which corresponds to the operator $D_{-\nu}.$
Consequently the integrals $K_{\pm\nu}P^{\pm}f$ taken over the sphere
$\Gamma^{R}$ (see the proof of Theorem \ref{Cauchyextt}) decrease when
$R\rightarrow\infty.$\medskip

Since $K_{\pm\nu}$ commute with $P^{\pm}$ we obtain that the integral
$K_{\overrightarrow{\alpha}}f(x)$ taken over the sphere $\Gamma^{R}$ also
decreases when $R\rightarrow\infty.$\medskip

In the case when $\alpha$ is a zero divisor the radiation condition
(\ref{radcondition}) becomes
\begin{equation}
\frac{ix}{\left|  x\right|  }f\overrightarrow{\alpha}=o\left(  \frac
{1}{\left|  x\right|  }\right)  ,\qquad\text{when }\left|  x\right|
\rightarrow\infty. \label{radcondzerodiv}%
\end{equation}
\medskip

Since the behavior of $\theta_{0}$ in (\ref{V0}) is of the type $O\left(
\frac{1}{\left|  x\right|  }\right)  $ when $\left|  x\right|  \rightarrow
\infty$ and since the expression $M^{\overrightarrow{\alpha}}V_{0}$ contains
the multiplication from the right-hand side by $\overrightarrow{\alpha}$, and
in this case $f\overrightarrow{\alpha}=o\left(  \frac{1}{\left|  x\right|
}\right)  $, it can be seen that the integral (\ref{kalfa}) taken over the
sphere $\Gamma^{R}$ decreases at infinity also.\medskip
\end{proof}

We now consider the Dirac operator $D_{\overrightarrow{\alpha}}$,
$\overrightarrow{\alpha}:=-(i\omega i_{1}+mi_{2})$. Note that $\overrightarrow
{\alpha}\in\frak{S}$ iff $\omega^{2}=m^{2}.$ In the following lines we will
obtain the radiation condition for the Dirac operator in its classical
representation using $\gamma$-matrices.\medskip

Let us denote
\begin{equation}
q:=A^{-1}f. \label{qdef}%
\end{equation}
According to (\ref{Dc}), if $f\in\ker D_{\overrightarrow{\alpha}}$ then
$q\in\ker\mathbb{D}_{\omega,m}$. Suppose that $f$ fulfills the radiation
condition (\ref{radcondition}) then we have
\[
\nu A\gamma_{1}\gamma_{2}\gamma_{3}A^{-1}f+A\gamma_{1}\gamma_{2}\gamma
_{3}A^{-1}\frac{ix}{\left|  x\right|  }f\overrightarrow{\alpha}=o\left(
\frac{1}{\left|  x\right|  }\right)  .
\]
\medskip

Let us denote
\[
f\overrightarrow{\alpha}=:F
\]
and consider the product
\[
xF=(x_{1}i_{1}+x_{2}i_{2}+x_{3}i_{3})F\text{.}\medskip
\]
Using properties 1.-3. of the transforms $A$ and $A^{-1}$ (Section 2.3), we
obtain
\[
xF=A\gamma_{1}\gamma_{2}\gamma_{3}\left(  \sum_{k=1}^{3}x_{k}\gamma
_{k}\right)  A^{-1}[F].
\]
\medskip Consequently, using the properties 4. and 5., we have
\[
xF(x)=A\gamma_{1}\gamma_{2}\gamma_{3}\left(  \sum_{k=1}^{3}x_{k}\gamma
_{k}\right)  A^{-1}[f(x)\cdot(-i\omega i_{1}-mi_{2})]=
\]%
\[
=A\gamma_{1}\gamma_{2}\gamma_{3}\left(  \sum_{k=1}^{3}x_{k}\gamma_{k}\right)
A^{-1}A\gamma_{1}\gamma_{2}\gamma_{3}\gamma_{0}A^{-1}f(x)\cdot(-i\omega)+
\]%
\[
+A\gamma_{1}\gamma_{2}\gamma_{3}\left(  \sum_{k=1}^{3}x_{k}\gamma_{k}\right)
A^{-1}A\gamma_{1}\gamma_{2}\gamma_{3}A^{-1}f(x)\cdot(-im)=
\]%

\[
=A\left(  \sum_{k=1}^{3}x_{k}\gamma_{k}\right)  \gamma_{0}A^{-1}%
f(x)\cdot(-i\omega)+A\left(  \sum_{k=1}^{3}x_{k}\gamma_{k}\right)
A^{-1}f(x)\cdot(-im).
\]
Using (\ref{qdef}) we can rewrite the previous equation as%

\[
xF\text{ }=A\left(  \sum_{k=1}^{3}x_{k}\gamma_{k}\right)  (-i\omega\gamma
_{0}-im)q.
\]
Thus (\ref{radcondition}) can be rewritten as follows
\[
\nu f+\frac{ix}{\left|  x\right|  }f\overrightarrow{\alpha}=\nu Aq+A\frac
{1}{\left|  x\right|  }\sum_{k=1}^{3}x_{k}\gamma_{k}(\omega\gamma
_{0}+m)q=o\left(  \frac{1}{\left|  x\right|  }\right)  ,
\]
and hence the radiation condition for the Dirac operator in its traditional
form is obtained in the following form
\begin{equation}
\nu q(x)-(\omega\gamma_{0}-m)\frac{\overrightarrow{x_{\gamma}}}{\left|
x\right|  }q(x)=o\left(  \frac{1}{\left|  x\right|  }\right)  ,\qquad\left|
x\right|  \rightarrow\infty\label{radcondclassic}%
\end{equation}
where $\overrightarrow{x_{\gamma}}:=\sum_{k=1}^{3}x_{k}\gamma_{k}$.\medskip

We introduce the following operator
\[
\mathbb{K}_{\omega,m}:=A^{-1}K_{\overrightarrow{\alpha}}A,
\]
where $\overrightarrow{\alpha}=-(i\omega i_{1}+mi_{2})$. Using this operator,
in \cite{OBBM} an analogue of the Cauchy integral formula for the Dirac
operator $\mathbb{D}_{\omega,m}$ in bounded domains was proved. Here with the
aid of the obtained radiation condition we obtain this fact for the unbounded domains.

\begin{theorem}
(Cauchy integral formula for $\mathbb{D}_{\omega,m}$ in the exterior domain
$\mathbb{R}^{3}\backslash\overline{\Omega}$) Let $q\in C^{1}(\mathbb{R}%
^{3}\backslash\overline{\Omega};\mathbb{H}(\mathbb{C}))\cap C(\mathbb{R}%
^{3}\backslash\Omega;\mathbb{H}(\mathbb{C}))$, $q\in\ker\mathbb{D}_{\omega
,m}(\mathbb{R}^{3}\backslash\overline{\Omega})$, and $q$ satisfies the
radiation condition (\ref{radcondclassic}). Then
\[
q(x)=\mathbb{K}_{\omega,m}[q](x),\qquad x\in(\mathbb{R}^{3}\backslash
\overline{\Omega}).
\]
\end{theorem}

The proof consists in the application of (\ref{Dc}) and Theorem
\ref{Cauchyextt}.

Acknowledgment: This work was partially supported by CONACYT, project 32424-E.


\medskip

\begin{thebibliography}{99}
\bibitem{QF}Bogoliubov N. N.and Shirkov D. V. \textit{Quantum fields}. Moscow:
Fizmatlit, 1993, 336 pp. (in Russian).

\bibitem {HFHE}G\"{u}rlebeck K. \textit{Hypercomplex factorization of the
Helmholtz equation}. Zeitschrift f\"{u}r Analysis und ihre Anwendungen, 1986,
v. 5, \#2, 125-131.

\bibitem {QCCP}G\"{u}rlebeck K. and Spr\"{o}ssig W. \textit{Quaternionic and
Clifford calculus for physicists and engineers.} John Wiley \& Sons, 1997.

\bibitem {OBBM}Kravchenko V. V. \textit{On a biquaternionic bag model.
}Zeitschrift f\"{u}r Analysis und ihre Anwendungen,1995, v. 14, \#1, 3-14.

\bibitem {ORBH}Kravchenko V. V. \textit{On the relation between holomorphic
biquaternionic functions and time-harmonic electromagnetic fields. }Deposited
in UkrINTEI, 29.12.1992, \#2073-Uk-92,18pp. (in Russian).

\bibitem {IRSM}Kravchenko V. V. and Shapiro M. V. \textit{Integral
representations for spatial models of mathematical physics}. Addison Wesley
Longman Ltd., Pitman Res., Notes in Math Series, v. 351, 1996.

\bibitem {CAME}McIntosh A. and Mitrea M. \textit{Clifford algebras and
Maxwell's equations in Lipschitz domains}. Mathematical Methods in the Applied
Sciences, 1999, v. 22, 1599-1620.

\bibitem {Muller}M\"{u}ller C. \textit{Grund probleme der mathematischen
Theorie elektromagnetischer Schwingungen. }Springer-Verlag, Berlin, 1957.

\bibitem {Silver}Silver S. \textit{Microwave antenna theory and design.
}M.I.T. Radiation Laboratory Series. Vol. 12, Mc Graw-Hill, New York, 1949.

\bibitem {Sommer}Sommerfeld A. \textit{Die Greensche Funktion der
Schwingungsgleichung}. Jber. Deutsch. Math. Verein. \textbf{21}, 309-353, 1912.

\bibitem {TDE}Thaller B. \textit{The Dirac equation}. Springer-Verlag, 1992.

\bibitem {EOMP}Vladimirov V. S. \textit{Equations of mathematical physics.
}Moscow: Nauka, 1984 (in Russian); Engl. transl. of the first edition: N.Y.:
Marcel Dekker, 1971.
\end{thebibliography}
\end{document}